\documentstyle[twocolumn,aps,prl,epsfig]{revtex}

\begin{document}
\twocolumn[\hsize\textwidth\columnwidth\hsize\csname@twocolumnfalse\endcsname

\title{Capillary to bulk crossover of nonequilibrium fluctuations in the
free diffusion of a near--critical binary liquid mixture.}
\author{ Pietro Cicuta$^\dag$, Alberto Vailati$^*$ and Marzio Giglio}
\address{Dipartimento di Fisica and Istituto Nazionale per
la Fisica della Materia, Universit\`a di Milano,\\ via Celoria 16,
20133 Milano, Italy.\\ $^\dag$ {current address: Department of
Physics, Cavendish Laboratory\\ Madingley Road, Cambridge CB3 0HE,
U.K.}\\ {$^*$\,E-mail address: vailati@fisica.unimi.it}}
\maketitle

\begin{abstract}
We have studied the nonequilibrium fluctuations occurring at the interface
between two miscible phases of a near-critical binary mixture during a free
diffusion process. The small-angle static scattered intensity is the
superposition of nonequilibrium contributions due to capillary waves and to
bulk fluctuations. A linearized hydrodynamics description of the
fluctuations allows us to isolate the two contributions, and to determine an
effective surface tension for the nonequilibrium interface. As the diffuse
interface thickness increases, we observe the cross-over of the
capillary-wave contribution to the bulk one.
\end{abstract}

%
%
%
%

%
%
%

%
%
%
%
\vskip 1cm OCIS: 240.6700, 290.5870 \vskip 2cm


\twocolumn]\narrowtext
\newpage
\section{Introduction}

In this work we focus on the investigation of interface and bulk
fluctuations in a near-critical binary liquid mixture undergoing a
free-diffusion process. free-diffusion processes have been recently shown to
give rise to giant nonequilibrium fluctuations \cite{nature,doriano}.
However, it is still not well understood what happens in the very early
stages of the diffusive remixing, when an effective surface tension should
be detectable \cite{petitjeans}. In this paper we use static light
scattering to study the crossover between the capillary-waves at the
interface and the bulk fluctuations, during the quick transient leading to
the onset of giant nonequilibrium fluctuations. We show that the static
light scattered from the diffusing sample is the superposition of a
contribution due to interface fluctuations and one due to bulk fluctuations.
By isolating the interface term we are able to recover the time evolution of
the nonequilibrium surface tension. We show that the surface tension
decreases very quickly, giving rise to a divergence of the amplitude of
capillary-waves. However, as the interface thickness increases due to
diffusion, a crossover from capillary to bulk fluctuations is observed: the
interface layer becomes thicker than the length scales associated with the
scattering wave vectors. Therefore, fluctuations do not displace this layer
as a whole any more. Instead, they take place inside the layer, which
behaves as a bulk phase. This prevents the divergence of the capillary-waves.

Similar free-diffusion experiments on critical fluids were performed on
isobutyric acid-water (IBW) by May and Maher \cite{may}, and on
cyclohexane-methanol (CM) and deuterated cyclohexane-methanol (CDM) by Vlad
and Maher \cite{vlad}. They used surface light scattering to recover the
relaxation time of fluctuations at the diffusing interface, and they applied
the capillary-waves dispersion relation to determine an effective surface
tension. They derived the time evolution of what they assumed to be an
effective surface tension of the interface. The time constant related to the
relaxation of the surface tension led to a value of the diffusion
coefficient which was many orders of magnitude smaller then the equilibrium
one (2 order of magnitude for CM, 3-5 for CDM and 7-8 for IBW). These
results were interpreted by assuming that gravity was responsible of the
slow dissolution of the interface. However, Maher and coworkers did not take
into account the presence of non-equilibrium bulk fluctuations at all. These
fluctuations turn out to be much slower than the capillary ones and this
might explain the slow relaxation time of the fluctuations they observed,
without having to resort to a slow-diffusion hypothesis.

\section{Experiment}

Several authors have tried to detect an effective surface tension between
two miscible fluids \cite{petitjeans,may,vlad,smith,joseph,mungall}. In
particular, some experiments have been performed by injecting a fluid into
the other one or by layering the fluid one on the top of each other \cite%
{petitjeans,smith,joseph}. However, it is very difficult to establish a flat
interface without generating disturbances. The use of a critical binary
liquid mixture has several advantages in this respect. A stable horizontal
equilibrium interface between two phases can be obtained by bringing the
mixture below its critical temperature. These two phases become completely
miscible simply by raising the temperature above the critical one, without
having to mechanically manipulate the sample. In this way disturbances in
the formation of the interface can be avoided, provided that some
precautions are taken to avoid convection. Moreover, working with a critical
mixture allows one to tune the time scales of fluctuations by changing the
distance from the critical point.

The experiment reported here has close connections with a previous
experiment on the transient behavior of a nonequilibrium interface in a near
critical binary mixture\cite{pietro}. This experiment was performed by
raising the temperature of a binary mixture from a value below $T_{c}$ to a
value still below it. In this way the system was separated into two phases
at the beginning and again at the end of the experiment, and we investigated
the transient behavior of the nonequilibrium interface and of the related
surface tension. In the experiment described here the temperature is raised
from below to above the critical one. In this way one starts with two phases
and ends with an homogeneous phase. The two concerns are how does the
interface evolve into a bulk phase and what happens to its surface tension.

The sample used is the binary mixture aniline-cyclohexane prepared at its
critical aniline weight fraction concentration c=0.47. A horizontal, 4.5 mm
thick, layer of the mixture is contained in a Rayleigh-Benard cell
consisting of two massive sapphire plates in thermal contact with two
Peltier elements controlled by a proportional-integral servo \cite%
{nature,soret}. The sapphire windows act at the same time as thermal plates
and as optical windows. The thermal stability of the cell is about 3 mK over
one week , the temperature being uniform within 2 mK across the sample. The
mixture is phase separated at 3 K below its critical temperature $T_{c}$, so
that two macroscopic phases are formed, separated by a horizontal interface.
The temperature is then suddenly increased to 1 K above $T_{c}$, so that the
two phases become completely miscible. During the temperature jump the
temperature of the upper plate is kept a few mK higher than that of the
lower one, so as to discourage convection. This thermal conditioning of the
sample has been thoroughly checked for spurious disturbances by using a very
sensitive shadowgraph technique. Scattered intensity distributions are
collected both during the initial equilibrium condition and during the
temperature increase by using a unique ultra low-angle static light
scattering machine described in detail elsewhere \cite{carpineti,ferri}. The
heart of this machine consists of a unique solid state sensor which is able
to collect light scattered within a two decades wave vector range, with a
dynamic range of about 6 decades. This sensor is a monolithic array of 31
photodiodes, having the shape of a quarter of an annulus. Each photodiode
collects the light scattered at a certain wave vector. The machine in the
configuration used for this experiment is able to collect the light
scattered in the wave vector range $80\,$cm$^{-1}<q<8000\,$cm$^{-1}$ in a
fraction of a second.

\section{Results and Discussion}

Prior to the temperature increase most of the light is scattered by
capillary-waves at the interface, the scattering due to the equilibrium bulk
phases being negligible. The intensity scattered by equilibrium
capillary-waves is well known to display the power-law behavior $I(q)\propto
1/(\sigma \,q^{2})$ \cite{mandelstamm}, where $\sigma $ is the surface
tension. The scattered intensity distribution in this condition is shown by
the plusses in Fig. 1, together with the best fit to the power law. As soon
as the temperature is raised, the scattered intensity increases at every
wave vector and a typical length scale appears at intermediate wave vectors.
The scattering data can be interpreted by following the guidelines outlined
in Ref. \cite{pietro} and summarized below. The concentration profile is
assumed to be continuous across the sample. The sample is modeled as the
superposition of an interface layer surrounded by two bulk layers. The
character of a layer depends on whether the typical length scale associated
with changes in the concentration gradient is respectively smaller or larger
than that associated with a fluctuation of wave vector $q$. Therefore a
certain layer of fluid can behave as an interface or as a bulk phase,
depending on the wave vector of the fluctuation perturbing it. Landau's
Fluctuating Hydrodynamics \cite{landau} can be used to describe the spectrum
of the fluctuations. By using the approach outlined in \cite{pre teo}, and
extended to nonequilibrium interfaces in \cite{pietro}, one finds that the
static contributions to the scattered intensities due to the interface and
bulk fluctuations are

\begin{equation}
I_{_{\text{int}}}(q)\propto \frac{\Delta c_{\text{int}}}{1+(q/q_{\text{cap}%
})^{2}}  \label{spettro int}
\end{equation}%
\qquad \qquad \qquad

\begin{equation}
I_{\text{bulk}}(q)\propto \frac{\Delta c_{\text{bulk}}}{1+(q/q_{\text{RO}%
})^{4}}  \label{spettro bulk}
\end{equation}

where%
\begin{equation}
q_{\text{cap}}=\left( \frac{\rho \beta g\Delta c_{\text{int}}}{\sigma }%
\right) ^{\frac{1}{2}}  \label{qcap}
\end{equation}

and

\begin{equation}
q_{\text{RO}}=\left( \frac{\beta g\nabla c}{\newline
\nu D}\right) ^{\frac{1}{4}}  \label{qro}
\end{equation}

In Eqs. (\ref{spettro int}-\ref{qro}) $\Delta c_{\text{int}}$ and $\Delta c_{%
\text{bulk}}$ represent the macroscopic concentration difference across the
interface and bulk layers, respectively, $g$ is the acceleration of gravity.
We assume that the concentration difference across the sample is small, so
that the fluid properties do not change much. Therefore, the mass density $%
\rho $, the kinematic viscosity $\nu $, the diffusion coefficient $D$ and $%
\beta =\rho ^{-1}(\partial \rho /\partial c)$ represent suitable average
values.

Equation (\ref{spettro int}) and (\ref{spettro bulk}) have basically the
same form: a power law at large wave vectors, and a gravitationally induced
saturation at smaller ones. Both capillary and bulk fluctuations are
generated by concentration fluctuations induced by velocity fluctuations due
to the presence of a macroscopic concentration gradient. Basically a parcel
of fluid gets displaced due to a thermal velocity fluctuation. Whenever this
displacement has a component parallel to a macroscopic concentration
gradient, a concentration fluctuation arises, as the parcel of fluid is
displaced into a region having a different concentration. The power-law
behavior at large wave vectors characterizes these velocity-induced
concentration fluctuations. In the case of an interface, the macroscopic
concentration changes across a length scale much smaller than those
associated with the scattering wave vectors. Therefore, a fluctuation
displaces this region of sharp concentration variation as a whole, and
relaxes back to equilibrium due to the capillary forces associated with the
interfacial displacement. This leads to the power-law exponent -2 in Eq. (%
\ref{spettro int}). In the case of a bulk phase, the macroscopic
concentration changes across a distance much larger than the length scales
associated with the wave vectors. In this way the fluctuations take place
'inside' the concentration gradient and relax back to equilibrium due to
diffusion. This mechanism leads to the exponent -4 of Eq. (\ref{spettro bulk}%
), typical of nonequilibrium fluctuations in a bulk phase \cite%
{kirk,ronis,law,schmitz,law2,pre teo,law3,segre1,segre2,li,soret,nature}.

The capillary and roll-off wave vectors $q_{\text{cap}}$ and $q_{\text{RO}}$
characterize the onset of the gravitational stabilization of the capillary
and bulk fluctuations, respectively. As outlined above, the interface and
bulk concentration fluctuations are brought about by the displacement of
parcels of fluid due to velocity fluctuations parallel to the macroscopic
gradient. As a consequence of the displacement, a buoyancy force acts on the
parcel, as its density is now different from that of the surrounding fluid.
Therefore gravity acts as a stabilizing mechanism against small wave vector
velocity fluctuations. This saturation is apparent for the late time data in
Fig. 1, while it falls outside the wave vector range for the early time data.

We assume that the interface and bulk phases scatter light independently, so
that the scattered intensity is a superposition of their contributions:

\begin{equation}
I(q,t)=I_{\text{int}}(q,t)+I_{\text{bulk}}(q,t)  \label{intensity}
\end{equation}

During a free-diffusion process the system is assumed to be infinite in the
vertical direction, and the concentration difference across the sample
height does not change in time. In practice the vertical extension of a
diffusion cell is finite, and the assumption is met only at the beginning of
a diffusion process, when the diffusive remixing involves the layers of
fluid near the interface at the mid-height of the cell. The typical time for
diffusion to occur across a cell of thickness $s$ is $\tau _{\text{diff}%
}\,=\,s^{2}/(\pi ^{2}D)$. For times shorter than $\tau _{\text{diff}}$ the
system can be assumed to be in a free-diffusion condition. Therefore, in the
free-diffusion regime the concentration differences $\Delta c_{\text{int}}$
and $\Delta c_{\text{bulk}}$ are not independent, as their sum corresponds
to the conserved concentration difference across the sample:

\begin{equation}
\Delta c(t)=\Delta c_{\text{int}}(t)+\Delta c_{\text{bulk}}(t)=\Delta c(t=0)
\label{deltac}
\end{equation}

This conservation is apparent in the late scattered-intensity distributions
of Fig. 1, collapsing onto the same curve at small wave vectors. According
to our model, also the earlier data in Fig. 1 should collapse onto the same
curve, but this occurs at wave vectors smaller than those accessible by our
experimental setup.

As pointed out above, the initial data of Fig. 1 display the $q^{-2}$
power-law behavior typical of capillary-waves. These data are not affected
by bulk fluctuations, since their amplitude is very small at equilibrium.
The $q^{-2}$ dependence is also apparent at large wave vectors in the
subsequent intensity distributions

This shows that an effective surface tension is present during the
nonequilibrium diffusive remixing, in agreement with the suggestion by some
authors that an effective surface tension should exist between miscible
fluids in the presence of a composition gradient \cite%
{petitjeans,may,vlad,smith,joseph,mungall,joseph2,maritan}. This surface
tension is usually assumed to be related to the concentration gradient by
the mean-field equilibrium Cahn-Hilliard relation \cite{cahnhilliard}

\begin{equation}
\sigma \propto \int (\frac{dc}{dz})^{2}dz  \label{cahnhilliard}
\end{equation}

As soon as the diffusive process is started, the concentration gradient at
the interface begins to drop. This determines the decrease of the effective
surface tension defined by Eq.(\ref{cahnhilliard}), and, according to Eq.(%
\ref{spettro int}), the increase of the intensity scattered at large wave
vectors. After about 70 s the power-law exponent of the large wave vector
scattered intensity gradually changes from 2 to 4. This corresponds to the
capillary to bulk cross-over of the fluctuations at the interface. According
to Eq. (\ref{cahnhilliard}) , the effective surface tension vanishes due to
the diffusive remixing, and this should determine a divergence of the light
scattered from capillary-waves. However, diffusion also determines the
increase of the typical length scale $\Lambda _{\text{int}}$ associated with
the concentration change at the interface: the concentration gradient at the
interface gets smaller and the interface thicker. For a given scattering
wave vector $q$, as soon as $\Lambda _{\text{int}}$ gets larger than $2\pi
/q $, the interface appears as a bulk phase and the power law exponent
changes accordingly. Therefore diffusion prevents the divergence of
capillary fluctuations as $\sigma \rightarrow 0$. Of course this is a $q$%
-dependent phenomenon: the first modes to undergo this transition are the
highest wave vectors.We were not able to observe this transition at
intermediate wave vectors, where the contribution of bulk fluctuations to
the scattered intensity dominates.

By using Eqs. (\ref{spettro int})-(\ref{deltac}) we are in the unique
position to fit the data in Fig. 1 to determine the time evolution of the
nonequilibrium effective surface tension, together with the concentration
difference at the interface. Best fits are represented by the continuous
lines in Fig. 1. The time evolution of $\Delta c_{\text{int}}$ and of the
surface tension are shown in Fig. 2. The initial value of the surface
tension derived from the equilibrium capillary-waves is a factor of two
smaller than the reference data of Atack and Rice, obtained by the
capillary-rise technique \cite{atack}. This is due to the smaller
sensitivity of the static surface light scattering technique, whose
measurements are affected by larger errors than the capillary rise ones. Our
data show unambiguously that the surface tension decreases by roughly two
orders of magnitude in about one minute. Results for the surface tension at
times larger than 80 s are not presented, because after this time the
capillary fluctuations cease to exist in our wave vector range, due to their
crossover to bulk fluctuations, as seen in the abrupt change of the large
wave vector power-law exponent in Fig. 1. The relaxation time of the surface
tension is strongly coupled with the time needed to raise the temperature of
the sample, which is of the order of 60 s. Therefore we have not attempted
to relate it to a typical diffusive time, as it would be very difficult to
isolate its contribution. Results for $\Delta c_{\text{int}}$ show that the
concentration difference gradually diminishes during the diffusion process.
This is due to the fact that as the concentration gradient diminishes in
layers of fluid at the boundary with the bulk phases, these layers cease to
belong to the interface, because the length scale associated with variations
in $\nabla c$ gets larger than those related with the scattering wave
vectors. Therefore the diminishing of $\Delta c_{\text{int}}$ is due to the
gradual cross-over of the interface to a bulk phase.

Our results show unambiguously that the effective surface tension between
two miscible phases of a near-critical binary mixture disappears very
quickly during the diffusion process. Our results also show that after about
80 s the determination of the surface tension from the spectrum of scattered
light is no longer feasible. This is due to the fact that the crossover from
capillary to bulk fluctuations prevents the determination of the surface
tension even at scattering angles as small as those achieved by our light
scattering setup.

Maher and coworkers in their experiments \cite{may,vlad} attempted to derive
values of the surface tension during the free-diffusion in near critical
binary mixtures for times as long as some hours. In view of the arguments
above, we feel that they improperly applied the capillary-wave description
to the bulk fluctuations in a diffuse interface, and this lead them to a
puzzling discrepancy between equilibrium and nonequilibrium values of the
diffusion coefficient. Their measurements were performed by means of surface
quasi-elastic light scattering. By assuming that the dynamics of the
fluctuations is controlled by capillary phenomena, one finds that the
relaxation time constant is \cite{webb}

\begin{equation}
\tau _{\text{cap}}=\frac{4\eta }{\sigma q}  \label{tau cap}
\end{equation}

where $\eta $ is the shear viscosity. However, in the presence of a diffuse
interface, the crossover from interface to bulk fluctuations occurs, and
this leads to the dispersion relation \cite{pre teo}

\begin{equation}
\tau _{\text{bulk}}=\frac{1}{Dq^{2}}\frac{1}{1+\left( \frac{q_{\text{RO}}}{q}%
\right) ^{4}}  \label{tau bulk}
\end{equation}

Although these fluctuations have the same origin as capillary-waves, namely
velocity fluctuations parallel to the concentration gradient, the transition
from a sharp to a diffuse interface drastically modifies both the static and
the dynamic structure factor of the fluctuations. We shall now compare the
fluctuations time scales as measured by May and Maher with those of bulk
fluctuations. From the nonequilibrium surface tension data from Ref. \cite%
{may} we recover the order of magnitude of the relaxation time of the
fluctuations observed by May and Maher. Data for the surface tension in Ref.
1 are scattered in the range $10^{-4}-10^{-3}$ dyne/cm for wave vectors in
the range $65\,$cm$^{-1}<q<345\,$cm$^{-1}$. By using Eq. (\ref{tau cap}) we
recover $\tau \approx 0.1\,$s$-6\,$s, where we have assumed $\eta \approx
10^{-2}$ poise. To compare these time scales with those associated with bulk
nonequilibrium fluctuations it is essential to estimate the roll-off wave
vector $q_{\text{RO}}$ defined in Eq. (\ref{qro}). By assuming the diffusive
behavior of the concentration gradient at the midheight of the cell $\nabla
c=\Delta c/(4\pi Dt)^{1/2}$ and by using the reference values $\beta =0.05$ (%
$\beta \approx \lbrack \rho _{\text{A}}-\rho _{\text{I}}]/\rho _{\text{A}}$,
where $\rho _{\text{A}}=1\,$g/cm$^{3}$ is the density of pure water and $%
\rho _{\text{I}}=0.95\,$g/cm$^{3}$ that of pure isobutyric acid), $\Delta
c=0.1$, $D=1.8\times 10^{-8}$cm$^{2}$/s \cite{may}, we find $q_{\text{RO}%
}\approx 900\,$cm$^{-1}-1100\,$cm$^{-1}$ at times ranging from 20 minutes up
to 120 minutes from the start of the diffusion process. Therefore, May and
Maher have operated at wave vectors smaller that the roll off one. By using
Eq. (\ref{tau bulk}) for the time scale of nonequilibrium bulk fluctuations
we recover $\tau _{\text{bulk}}\approx 0.14\,$s$-10\,$s in the wave vector
range and time windows used by May and Maher. The agreement between the time
scales observed by them and those predicted by our model is more than good,
also considered that our description removes the puzzling discrepancy
between equilibrium and nonequilibrium values of the diffusion coefficient
reported by May and Maher.

The data presented in Ref. \cite{vlad} are of better quality than those in
Ref. \cite{may}, and are therefore suitable for a more quantitative check of
the time scales. However the authors did not to report the wave vectors at
which measurements were performed and therefore quantitative estimates are
difficult.

\section{Conclusions}

We have shown that the transient surface tension between two miscible phases
of a binary liquid mixture decreases to vanishingly small values in a very
short time. However the divergence of the associated capillary-waves is
prevented due to their crossover to nonequilibrium bulk fluctuations as the
interface gets thicker. We think that the fluctuations observed by May and
Maher, and by Vlad and Maher, are not related to surface tension effects at
all. The time scales of the fluctuations observed by them are fully
compatible with those predicted for nonequilibrium fluctuations in a diffuse
gradient. There is no ultra-slow macroscopic diffusion to speak of, and this
could be checked by means of traditional interferometric techniques \cite%
{tyrrell}.

This work was supported by the Italian Space Agency (ASI).

%
%
%
%


\begin{figure}[h]
\epsfig{file=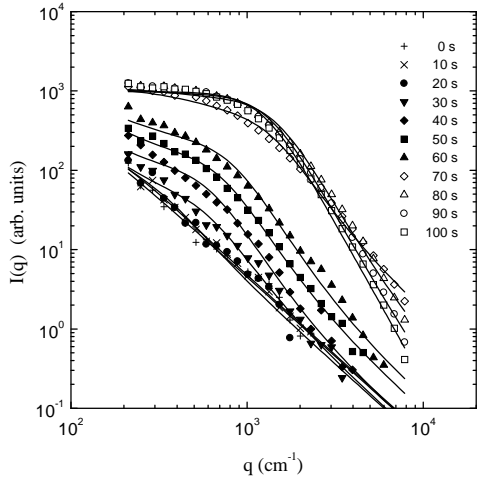,width=8cm}
\caption{Scattered intensity
plotted vs. scattered wave vector $q$ at different times. The
initial dataset~($+$) is the intensity scattered by the initial
equilibrium interface. Time is measured from the start of the
remixing process. The figure shows the early stages of the
remixing process, when the scattered intensity is increasing with
time. The solid lines
represent the best fit of the experimental data with Eqs.~(\ref{spettro int}-%
\ref{deltac}).}
\label{F1}
\end{figure}
\begin{figure}[h]
\epsfig{file=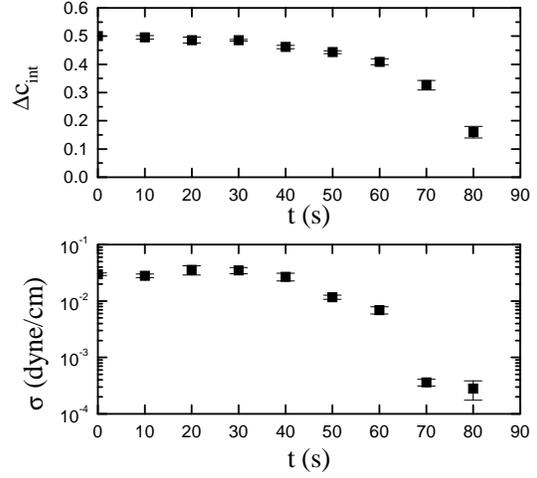,width=8cm}
\caption{ Time evolution of the concentration difference across the
interface, (a), and of the interfacial surface tension, (b). }
\label{F2}
\end{figure}


\begin{references}
\bibitem{nature} A. Vailati and M. Giglio, ''Giant fluctuations in a free
diffusion process'', Nature {\bf {390}}, 262-265 (1997).

\bibitem{doriano} D. Brogioli, A. Vailati, and \ M. Giglio, ''Universal
behavior of nonequilibrium fluctuations in free diffusion processes'', Phys.
Rev. E 61, R1-R4 (2000).

\bibitem{petitjeans} P. Petitjeans, ''Une tension de surface pour les
fluides miscibles'', C. R. Acad. Sci. Paris 322, 673-679 (1996).

\bibitem{may} S. E. May and J. V. Maher, ''Capillary-wave relaxation for a
meniscus between miscible liquids'', Phys. Rev. Lett. 67, 2013-2016 (1991).

\bibitem{vlad} D. H. Vlad and J.V. Maher, ''Dissolving interfaces in the
presence of gravity'', Phys. Rev. E 59, 476-478 (1999).

\bibitem{smith} P. G. Smith, M. Van den Ven, and S. G. Mason, ''The
transient interfacial tension between two miscible fluids'', J. Colloid
Interface Sci. 80, 302-303 (1981).

\bibitem{joseph} D. D. Joseph, '' Fluid dynamics of two miscible liquids
with diffusion and gradient stresses'', Eur. J. Mech., B/Fluids 9, 565-596
(1990).

\bibitem{mungall} J. E. Mungall, ''Interfacial tension in miscible two-fluid
systems with linear viscoelastic rheology'', Phys. Rev. Lett. 73, 288-291
(1994).

\bibitem{pietro} P. Cicuta, A. Vailati, and M. Giglio, ''Equilibrium and
nonequilibrium fluctuations at the interface between two fluid phases'',
Phys. Rev. E 62, 4290-4926 (2000).

\bibitem{soret} A.Vailati and M.Giglio, ''q divergence of nonequilibrium
fluctuations and its gravity induced frustration in a temperature stressed
liquid mixture'', Phys. Rev. Lett. 77, 1484-1487 (1996).

\bibitem{carpineti} M. Carpineti, F. Ferri, M. Giglio, E. Paganini, and U.
Perini, \ ''Salt-induced fast aggregation of polystyrene latex'', Phys. Rev.
A 42, 7347-7354 (1990).

\bibitem{ferri} A. Bassini, S. Musazzi, E. Paganini, U. Perini, and F.
Ferri, \ ''Self-aligning optical particle sizer for the monitoring of
particle growth processes in industrial plants'', Rev. Sci. Instrum. 69,
2484-2494 (1998).

\bibitem{mandelstamm} L. I. Mandelstamm, ''Uber \ die Rauhigkeit freier
Flussigkeitsoberflachen'', Ann. der Physik 41, 609-624 (1913).

\bibitem{landau} L. D. Landau, and E. M. Lifshifz, Fluid Mechanics
(Pergamon, New York, 1959).

\bibitem{pre teo} A.Vailati and M.Giglio, ''Nonequilibrium fluctuations in
time-dependent diffusion processes'', Phys. Rev. E 58, 4361-4371 (1998).

\bibitem{kirk} T. R. Kirkpatrick, E. G. D. Cohen, and J. R. Dorfman,
''Fluctuations in a nonequilibrium steady state. I. Small gradients'',
Phys.Rev.A 26, 995-972 (1982).

\bibitem{ronis} D. Ronis, and I. Procaccia, ''Nonlinear resonant coupling
between shear and heat fluctuations in fluids far from equilibrium'', Phys.
Rev.A 26, 1812-1815 (1982).

\bibitem{law} B. M. Law and J. C. Nieuwoudt, ''Noncritical liquid mixtures
far from equilibrium: the Rayleigh line'', Phys. Rev. A 40, 3880-3885 (1989).

\bibitem{schmitz} R. Schmitz and E. G. D. Cohen, ''Fluctuations in a fluid
under a stationary heat flux II. Slow part of the correlation matrix'', J.
Stat. Phys. 40, 431-482 (1985).

\bibitem{law2} B. M. Law and J. V. Sengers, ''Fluctuations in fluids out of
thermal equilibrium'', J. Stat. Phys. 57, 531-547 (1989).

\bibitem{law3} B. M. Law, R. W. Gammon and J. V. Sengers, ''Light-scattering
observations of long-range correlations in a nonequilibrium liquid'', Phys.
Rev. Lett. 60, 1554-1557 (1988).

\bibitem{segre1} P. N. Segre, R. W. Gammon, J. W. Sengers and B. M. Law,
''Rayleigh scattering in a liquid far from thermal equilibrium'', Phys. Rev.
A 45, 714-724 (1992).

\bibitem{segre2} P. N. Segre, R. W. Gammon and J. V.
Sengers,''Light-scattering measurements of nonequilibrium fluctuations in a
liquid mixture'', Phys. Rev. E 47, 1026-1034 (1993).

\bibitem{li} W. B. Li, P. N. Segre, R. W. Gammon, and J. V. Sengers,
''Small-angle Rayleigh scattering from nonequilibrium fluctuations in
liquids and liquid mixtures'', Physica A 204, 399-435 (1994).

\bibitem{joseph2} G. P. Galdi, D. D. Joseph, L. Preziosi, and S. Rionero,
''Mathematical problems for miscible, incompressible fluids with Korteweg
stresses'', Eur. J. Mech.,B/Fluids 10, 253-267 (1991).

\bibitem{maritan} W. J. Ma, P. Keblinski, A. Maritan, J. Koplik, and R.
Banavar, '' Dynamical relaxation of the surface tension of miscible
phases'', Phys. Rev. Lett 71, 3465-3468 (1993).

\bibitem{cahnhilliard} J. W. Cahn and J. E. Hilliard, ''Free energy of a
nonuniform system. Interfacial free energy'', J. Chem. Phys. 28, 258-267
(1958).

\bibitem{atack} D. Atack and O. K. Rice, ''The interfacial tension and other
properties of the cyclohexane + anyline system near the critical solution
temperature'', Discuss. Faraday Soc. 15, 210-218 (1953).

\bibitem{webb} J. S. Huang and W. W. Webb, ''Viscous damping of thermal
excitations on the interface of critical fluid mixtures'', Phys. Rev. Lett.
23, 160-163 (1969).

\bibitem{tyrrell} H. J. V. Tyrrell, Diffusion and Heat Flow in Liquids
(Butterworths, London, 1961).

\end{references}
\end{document}